\documentclass[letter,twocolumn]{jpsj3}
\usepackage{graphicx} 
\usepackage{dcolumn} 
\usepackage{latexsym}
\usepackage{txfonts}
\usepackage{color}
\begin{document} 
 
\title{ 
Rotation of Triangular Vortex Lattice 
in the Two-Band Superconductor ${\rm MgB_2}$ 
} 
\author{
Tomoya \surname{Hirano},  
Kenta \surname{Takamori},  
Masanori \surname{Ichioka}, and 
%~\thanks{E-mail address: ichioka@cc.okayama-u.ac.jp}}
Kazushige  \surname{Machida}
}

\inst{
Department of Physics, Okayama University, 
Okayama 700-8530, Japan}  

\date{\today}

\abst{
To identify the contributions of the multiband nature and the anisotropy of 
a microscopic electronic structure to a macroscopic vortex lattice morphology, 
we develop a method based on the Eilenberger theory near $H_{\rm c2}$
combined with the first-principles band calculation 
to estimate the stable vortex lattice configuration.
For a typical two-band superconductor ${\rm MgB_2}$, 
successive transitions of vortex lattice orientation 
that have been observed recently by small angle neutron scattering 
[Das, {\it et al.}: Phys. Rev. Lett. {\bf 108} (2012) 167001]
are explained 
by the characteristic field-dependence of two-band superconductivity and 
the competition of sixfold anisotropy 
between the $\sigma$- and $\pi$-bands. 
The reentrant transition at low temperature reflects 
the Fermi velocity anisotropy of the $\sigma$-band. 
} 

\kword{
vortex lattice morphology, 
two-band superconductivity, 
${\rm MgB_2}$, 
first-principles band calculation, 
quasi-classical Eilenberger theory} 
 
% \pacs{74.25.Uv, 74.25.Op, 74.20.Pq, 74.70.Ad} 
%74.25.Uv	Vortex phases (includes vortex lattices, vortex liquids, and vortex glasses)  
%74.25.Op	Mixed states, critical fields, and surface sheaths
%74.20.Pq	Electronic structure calculations (for methods of electronic structure calculations, see 71.15.-m)
%74.25.Ha	Magnetic properties including vortex structures and related phenomena (for vortices, magnetic bubbles, and magnetic domain structure, see 75.70.Kw)
%74.70.Ad	Metals; alloys and binary compounds (including A15, MgB2, etc.)
%71.15.-m	Methods of electronic structure calculations (see also 31.15.-p Calculations and mathematical techniques in atomic and molecular physics; for electronic structure calculations of superconducting materials, see 74.20.Pq)
%71.15.Ap	Basis sets (LCAO, plane-wave, APW, etc.) and related methodology (scattering methods, ASA, linearized methods, etc.)
%71.15.Dx	Computational methodology (Brillouin zone sampling, iterative diagonalization, pseudopotential construction)
%71.15.Mb	Density functional theory, local density approximation, gradient and other corrections

\maketitle 
%%%%%%%%%%%%%%%%%%%%%%%%%%%%%%%%%%%%%%%%%%%%%%%%%%%%%%%%%%%%%%%% 
%\section{introduction} 

Since vortices are self-forming objects 
in mixed states of type-II superconductors under applied magnetic fields, 
the properties of the vortex states give us valuable information 
on the characteristics of each superconductor.  
Among them, the vortex lattice morphology, 
depending on the temperature $T$ and magnetic field $H$,   
is closely related to the anisotropy of the superconductivity. 
For example, reflecting the fourfold symmetric structure 
of a superconducting gap or the Fermi velocity in momentum space, 
the vortex lattice shows a gradual transformation  
from a triangular vortex lattice at low fields to a square lattice 
at high fields~\cite{Suzuki}.   
The anomalous vortex lattice transformation in a superconductor Nb, 
including a scalene triangular lattice, 
is considered to be a result of the anisotropy of the Fermi surface 
structure~\cite{Laver,AdachiNb,Takanaka}.  
For a reliable theoretical estimation of 
the vortex lattice morphology, 
we need microscopic calculations such as the use of the Eilenberger theory, 
beyond the phenomenological ones of 
the extended Ginzburg-Landau (GL) theory~\cite{Takanaka,Zhitomirsky}  
or the nonlocal London theory~\cite{Kogan}.   
Furthermore, to accurately 
consider the anisotropy of the Fermi surface structure, 
it is preferable that 
the results of first-principles band calculation for electronic states, 
such as density functional theory (DFT), are combined in the estimation 
of a stable vortex lattice. 
This theoretical approach was carried out in a study of 
the vortex lattice morphology in Nb~\cite{AdachiNb}.   
The success encouraged us to further extend the vortex physics 
along this line.

\begin{figure}[tb] 
\begin{center}
\includegraphics[width=8.5cm]{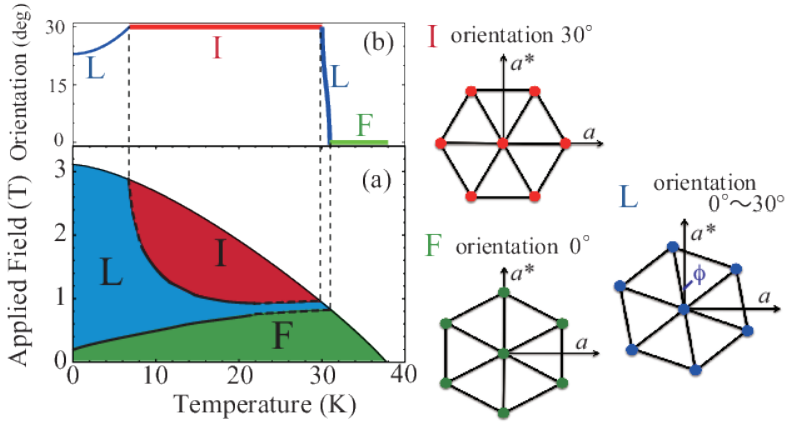} 
\vspace{-0.3cm}
\end{center}
 \caption{ 
(Color online)  
(a) Schematic phase diagram of vortex lattice  morphology 
as a function of $H$ and $T$, 
obtained by the SANS experiment~\cite{Das}. 
The phase boundary near $H_{\rm c2}$ is extrapolated.   
Triangular lattices in the right panel 
show the orientation $\phi$ of each-phase. 
(b) The $T$ dependence of $\phi$ is schematically presented along $H_{\rm c2}$.  
} 
\label{fig:exp} 
\end{figure}

The properties of a superconductor ${\rm MgB_2}$ 
as a typical multiband superconductor 
also attract much attention.  
The superconductivity in ${\rm MgB_2}$ 
consists of the $\sigma$-band with a large superconducting gap 
and the $\pi$-band with a small gap~\cite{Gonnelli,Schmidt}. 
In the vortex state, there are some interesting phenomena 
reflecting the multiband superconductivity, 
such as a rapid increase 
in the $H$-dependence of low-temperature specific heat $C(H)=\gamma(H)T$ 
at low fields,~\cite{Yang,Bouquet,NakaiJPSJ,Koshelev,IchiokaMgB2} 
and the vortex clustering of the so-called type 1.5 superconductivity 
at low fields~\cite{Moshchalkov}.   
Also, an interesting vortex lattice morphology has recently 
been reported in 
${\rm MgB_2}$ for $H \parallel c$~\cite{Cubitt,Das}. 
Since the crystal lattice is hexagonal, 
a triangular vortex lattice is formed in ${\rm MgB_2}$. 
As schematically shown in Fig. \ref{fig:exp}, from 
small angle neutron scattering (SANS) experiments~\cite{Das}, 
the orientation of the triangular vortex lattice is 
$\phi =0^\circ$ (${\rm F}$-phase) at low fields, 
and $\phi=30^\circ$ (${\rm I}$-phase) at high fields, 
where $\phi$ is the angle measured from the $a^\ast$-axis in real space 
($a$-axis in reciprocal space) of the underlying hexagonal crystal. 
In the intermediate fields, the orientation rotates gradually between 
$0^\circ < \phi < 30^\circ$ (${\rm L}$-phase).
However, at low-$T$, the high-field ${\rm I}$-phase reported 
in a previous work~\cite{Cubitt} has been confirmed as a metastable state in 
a more recent work~\cite{Das}. 
The true stable state at low-$T$ and high fields is found to be 
${\rm L}$-phase, as shown in Fig. \ref{fig:exp}. 
Therefore, a stable vortex lattice shows the successive transition 
${\rm L} \rightarrow {\rm I} \rightarrow {\rm L} \rightarrow {\rm F}$ 
along $H_{\rm c2}$ upon raising $T$ from $T=0$ to 
the transition temperature $T_{\rm c}$.
In a previous study based on phenomenological GL theory~\cite{Zhitomirsky},  
the transition ${\rm I} \rightarrow {\rm L} \rightarrow {\rm F}$ was 
considered as a result of competition in anisotropies between 
the $\sigma$-band and the $\pi$-band in two-band superconductivity. 
However, the reentrant transition ${\rm L} \rightarrow {\rm I}$ 
at low-$T$ has not yet been explained theoretically. 
We note that the original GL theory is valid only near $T_{\rm c}$.  
To discuss the vortex morphology of ${\rm MgB_2}$, 
we have to consider the 6th- and 12th-order derivative terms for 
gradient terms of the GL equation.
The nonlocal correction terms of the $2n$-th order derivative 
are on the order of $(1-T/T_{\rm c})^{n-1}$. 
Thus, it is difficult in the framework of the GL theory 
to discuss the vortex morphology in the low-temperature region, 
because we need further higher-order derivative correction terms. 
On the other hand, since the Eilenberger theory can be 
applied to all $T$ regions, we can obtain 
a reliable estimation for a vortex structure including that 
in a low-$T$ region. 
The purpose of this study is to investigate the successive transition of 
${\rm L} \rightarrow {\rm I} \rightarrow {\rm L} \rightarrow {\rm F}$ 
along $H_{\rm c2}$, 
on the basis of the Eilenberger theory of two-band superconductors 
combined with DFT calculation~\cite{Wien2k} 
to consider the Fermi surface of ${\rm MgB_2}$. 

%%%%%%%%%%%%%%%%%%%%%%%%%%%%%%%%%%%%%%%% 
%%%%%%%%%%%%%%%%%%%%%%%%%%%%%%%%%%%%%%%% 

First, we study the sixfold anisotropy of the Fermi surface structure 
in ${\rm MgB_2}$. 
The electronic state of ${\rm MgB_2}$ is evaluated by 
DFT calculation~\cite{Wien2k} using local density approximation and 
the lattice constants $a=b=3.083 {\rm \AA}$, 
and $c=3.521 {\rm \AA}$~\cite{Wan}. 
We chose the muffin-tin radii $R_{\rm MT}$ of 2.35 and 1.67 a.u., respectively, 
and use a plane-wave cutoff $R_{\rm MT,min}K_{\rm max}=9.0$.
The first Brillouin zone is divided into $120 \times 120 \times 90$ grids.
By the tetrahedron method~\cite{Blochl} 
for determing patches of the Fermi surface from the grids, we estimate 
the Fermi wave number ${\bf k}_i$, 
Fermi velocity ${\bf v}_i=(v_{i,x},v_{i,y},v_{i,z})$,   
and Fermi surface DOS $D_i(\epsilon_{\rm F})$ 
on the $i$-th patch of the Fermi surface. 

\begin{figure}[tb] 
\begin{center}
\includegraphics[width=8.5cm]{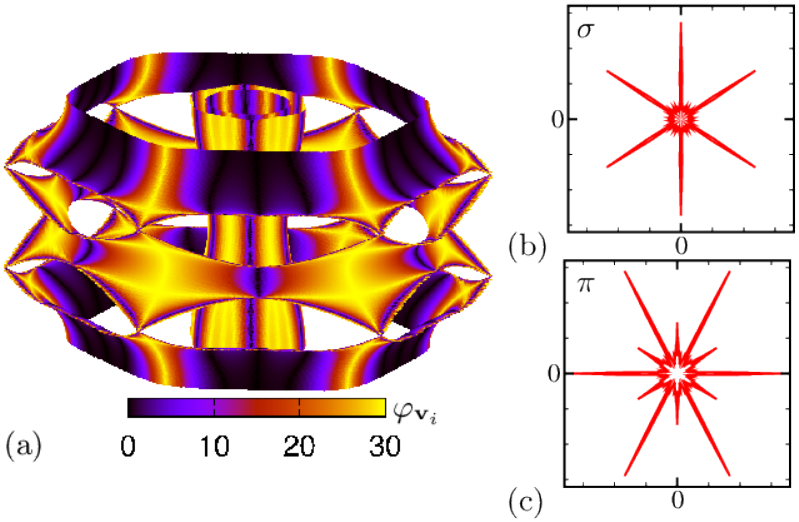} 
\vspace{-0.3cm}
\end{center}
 \caption{ 
(Color online) 
(a) 
Fermi surfaces of ${\rm MgB_2}$ obtained in our calculation. 
The two cylinders in the center are the $\sigma$-band's Fermi surface, 
and the outside rings are the $\pi$-band's Fermi surfaces. 
The color on the surface indicates the orientation $\varphi_{{\bf v}_i}$ 
(mod $30^\circ$) of the Fermi velocity. 
(b) $\varphi_v$-resolved Fermi surface DOS 
$D_\sigma(\varphi_v)$ for the $\sigma$-band. 
(c) $D_\pi(\varphi_v)$ for the $\pi$-band. 
In (b) and (c),  
we plot lines of 
$(D_m(\varphi_v)\cos\varphi_v, D_m(\varphi_v)\sin\varphi_v)$ 
for $0 \le \varphi_v \le 360^\circ$ ($m=\sigma,\pi$), 
so that the distances from the center to the lines indicate $D_m(\varphi_v)$.  
} 
\label{fig:DFT} 
\end{figure}

As shown in Fig. \ref{fig:DFT}(a), 
our calculation reproduces the well-known 
Fermi surfaces of the $\sigma$-band and the $\pi$-band~\cite{Kortus}. 
For later discussions on the reasons for the vortex lattice 
morphology in ${\rm MgB_2}$, 
we study the anisotropy of the Fermi velocity 
projected within the $xy$ plane 
perpendicular to the applied field along the $z$-direction. 
The color on the Fermi surfaces in Fig. \ref{fig:DFT}(a) 
shows the orientation angle $\varphi_{{\bf v}_i}$ of the Fermi velocity 
defined as $\tan \varphi_{{\bf v}_i}=v_{i,y}/v_{i,x}$. 
Figures \ref{fig:DFT}(b) and \ref{fig:DFT}(c) show 
the $\varphi_v$-resolved Fermi surface DOS defined as 
\begin{eqnarray}
D_m(\varphi_v)=\langle \delta(\varphi_v-\varphi_{{\bf v}_i})  \rangle_m , 
\end{eqnarray}
for each band ($m=\sigma,\pi$),  
where the 
band-resolved Fermi surface average is 
$\langle \cdots \rangle_m=\sum_{i\in m} D_i(\epsilon_{\rm F}) (\cdots)$.  
As shown in Fig. \ref{fig:DFT}(b), 
the Fermi surfaces of the $\sigma$-band are  
two hexagonal cylinders rather than circular ones. 
Most of the Fermi velocity on the flat surface is oriented to 
$30^\circ$ or equivalent directions. 
The Fermi surface of the $\pi$-band is also flat, and 
most of the Fermi velocity on the flat surface is oriented to 
$0^\circ$  or equivalent directions, as shown in Fig. \ref{fig:DFT}(c). 
Therefore, the $D_m(\varphi_v)$ values of the Fermi surface are different 
between the two bands, inducing a competition between the ${\rm F}$ and 
${\rm I}$ phases for the vortex lattice orientation, 
as will be discussed later. 
The ratio of the Fermi surface DOS of 
the $\pi$-band to the $\sigma$-band 
is calculated to be $N_\pi/N_\sigma =1.36$, 
indicating that the $\pi$-band contribution is larger. 

 %%%%%%%%%%%%%%%%%%%%%%%%%%%%%%%%%%%%%%%% 
%%%%%%%%%%%%%%%%%%%%%%%%%%%%%%%%%%%%%%%% 

Next, we explain our formulation for estimating  
the stable vortex lattice configuration, 
which is extended to two-band superconductors in this work. 
To shorten our calculation within a reasonable computational time 
even when we consider the detailed structure of three-dimensional 
Fermi surfaces, 
our study of the stable vortex lattice is restricted near $H_{\rm c2}$, 
as a first approach to this problem. 
We assume an isotropic s-wave superconducting gap in each band, 
and a clean limit.  

Quasi-classical Green's functions 
$g( \omega_n, {\bf k}_i,{\bf r})$,  
$f( \omega_n, {\bf k}_i,{\bf r})$, and  
$f^\dagger( \omega_n, {\bf k}_i,{\bf r})$ 
with a Matsubara frequency $\omega_n$ are 
defined at ${\bf k}_i$ on each Fermi surface. 
They are calculated by the Eilenberger 
equation~\cite{Eilenberger,IchiokaMgB2} 
\begin{eqnarray} && 
\left\{ \omega_n  
+{\bf v}_i \cdot\left(\nabla+{\rm i}{\bf A} \right)\right\} 
f( \omega_n, {\bf k}_i,{\bf r}) 
=\Delta_m({\bf r}) g( \omega_n, {\bf k}_i,{\bf r})  \qquad
\label{eq:Eil}
\end{eqnarray}  
and the equivalent equation for $f^\dagger$, 
where $g=(1-ff^\dagger)^{1/2}$ 
and $\bf A$ is the vector potential. 
Throughout this letter, we use Eilenberger units,~\cite{Klein} 
i.e.,   
energy, temperature, and length are in units of $\pi k_{\rm B} T_{\rm c}$, 
$T_{\rm c}$, and $\xi_0=\hbar v_{{\rm F}}/2\pi k_{\rm B} T_{\rm c}$, respectively. 
The self-consistent gap equation for the pair potential 
$\Delta_m({\bf r})$ of $m$-bands ($m=\sigma$, $\pi$) is given by
 \begin{eqnarray}
\left(\begin{array}{c} 
 \Delta_\sigma({\bf r}) \\ \Delta_\pi({\bf r})  \\ 
\end{array}\right)
= 
\hat{V} 
T \sum_{\omega_n} 
\left(\begin{array}{c} 
\langle f(\omega_n,{\bf k}_i,{\bf r}) \rangle_\sigma \\ 
\langle f(\omega_n,{\bf k}_i,{\bf r})  \rangle_\pi \\ 
\end{array}\right). 
\label{eq:gap}
\end{eqnarray}  
As for the interaction constants of superconductivity, 
\begin{eqnarray}
\hat{V}=
\left(\begin{array}{cc} 
 V_{\sigma,\sigma} & V_{\sigma,\pi} \\ V_{\sigma,\pi} & V_{\pi,\pi} \\ 
\end{array}\right), 
\end{eqnarray} 
where 
$V_{\sigma,\sigma}$ ($V_{\pi,\pi}$) is the pairing interaction 
in the $\sigma$- ($\pi$-) band, 
and $V_{\sigma,\pi}$ is the Cooper pair transfer between two bands. 
We have to select these parameters 
under the condition to reproduce the gap ratio 
$\Delta_\pi/\Delta_\sigma = \frac{1}{4}\sim\frac{1}{2}$ 
observed in experiments~\cite{Schmidt}.   
Previous first-principles band calculation~\cite{Golubov} estimated 
(i) $V_{\sigma,\pi}=0.15V_{\sigma,\sigma}$ and $V_{\pi,\pi}=0.32V_{\sigma,\sigma}$.  
These are used in our calculations.  
We determine $V_{\sigma,\sigma}$ from the gap equation at $T=T_{\rm c}$. 
 
We expand  $\Delta_m({\bf r})$ and 
$f(\omega_n, {\bf k}_i,{\bf r})$ in terms of the Landau-Bloch function 
$\psi_{N, {\bf q}}({\bf r})$~\cite{KitaPRB,KitaJPSJ}, 
whose the $N$-th coefficients of the Landau level 
are $\Delta_{m,N}$ and $f_N(\omega_n, {\bf k}_i)$. 
The vortex lattice configuration is determined by the parameters 
of $\psi_{N, {\bf q}}({\bf r})$, and the relative angle 
of the crystal coordinate and vortex coordinate. 
Along $H_{c2}$ where $g \rightarrow {\rm sgn}{\omega_n}$, 
the Eilenberger equation (\ref{eq:Eil}) is reduced to~\cite{KitaPRB,KitaJPSJ} 
\begin{eqnarray}
f_N(\omega_n, {\bf k}_{i \in m})
=\sum_{M} ({\cal M}^{-1})_{N,M} \Delta_{m,M}
\label{eq:fN} 
\end{eqnarray} 
with 
${\cal M}_{N,M}
=|\omega_n|\delta_{N,M}
+\sqrt{N+1} b ^\ast \delta_{N,M-1} -\sqrt{N} b  \delta_{N,M+1} 
$ 
and 
$ b =(v_{i,x}+{\rm i}v_{i,y}) {\rm sgn}(\omega_n) \sqrt{H} / (2 \sqrt{2})$. 
The Fermi velocities $v_{i,x}$ and $v_{i,y}$ at ${\bf k}_i$, appearing in $b$, 
determine the ${\bf k}_i$ dependence of $f_N$. 
Therefore, at $B=H_{\rm c2}$, the gap equation (\ref{eq:gap}) is reduced to 
 \begin{eqnarray}
\left(\begin{array}{c} 
 \Delta_{\sigma,N} \\ \Delta_{\pi,N}  \\ 
\end{array}\right)
= 
\hat{V} 
T \sum_{\omega_n} 
\sum_M 
\left(\begin{array}{c} 
\langle ({\cal M}^{-1})_{N,M} \rangle_\sigma  \Delta_{\sigma,M} \\ 
\langle ({\cal M}^{-1})_{N,M} \rangle_\pi     \Delta_{\pi,M} \\ 
\end{array}\right). 
\label{eq:gapM}
\end{eqnarray}  

In the Eilenberger theory, 
the free energy in a two-band superconductor is given by 
$F=\kappa^2 B^2 +F_{\rm c1}+F_{\rm c2}$ with the GL parameter $\kappa$ and 
\begin{eqnarray} && 
F_{\rm c1}=\sum_{n,m=\sigma,\pi} 
\overline{\Delta^\ast_n({\bf r}) (\hat{V}^{-1})_{n,m} \Delta_n({\bf r}) }
,  
\\ && 
F_{\rm c2}= -T\sum_{\omega_n} \sum_m \overline{\langle I\rangle_m},  
\  
\end{eqnarray}
where $\overline{(\cdots)}$ indicates the spatial average and  
$I=(g-{\rm sgn}\omega_n)
[2\omega_n+{\bf v}_i \cdot(\nabla\ln(f/f^{\dagger})
 +2{\rm i}{\bf A})]+ f\Delta^{\ast}+\Delta f^{\dagger}$.
Using the gap equation (\ref{eq:gap}), we obtain 
\begin{eqnarray}
F_{\rm c1}=\frac{T}{2}\sum_{\omega_n} \sum_m \overline{ 
(\Delta_m \langle f^\dagger \rangle_m + \Delta_m^\ast \langle f \rangle_m )}. 
\end{eqnarray}
Since the condensation energy part $F_{\rm c1}+F_{\rm c2}$ is 
a simple summation of each band contribution,  
the derivation of free energy near $H_{\rm c2}$ can be performed 
as in a single band case~\cite{Nakai,Abrikosov,deGennes}, as follows.   
We substitute $g\sim 1-ff^{\dagger}/2-(ff^{\dagger})^2/8$ in $F$ 
and set ${\bf A}={\bf A}_{\rm c2}+{\bf A}_1$, 
where $\nabla\times{\bf A}_{\rm c2}={\bf H}_{\rm c2}$, 
$\nabla\times{\bf A}_1={\bf h}_1$, and 
${\bf h}_1={\bf H}-{\bf H}_{\rm c2}+{\bf h}_s$ 
for an applied field $H$ near $H_{\rm c2}$. 
Thus, the Abrikosov identity~\cite{Abrikosov} is derived as 
\begin{eqnarray} 
T \sum_{\omega_n >0} 
\sum_m 
\overline{ \langle f f^\dagger (\Delta_m f^\dagger+\Delta_m^\ast f)\rangle_m} 
- 4 \kappa^2 \overline{{\bf h}_1 \cdot {\bf h}_s}=0 , 
\end{eqnarray}
where ${\bf h}_s$ is the magnetic field induced by supercurrent, and 
\begin{eqnarray}
\nabla \times {\bf h}_s 
= \frac{T}{\kappa^2} \sum_{\omega_n>0} \sum_m 
  \langle {\bf v} f f^\dagger \rangle_m . 
\end{eqnarray} 
Finally, we obtain the free energy 
at magnetic induction $B=H+\bar{h}_s$ near $H_{c2}$ as~\cite{Nakai}
\begin{eqnarray} 
\frac{F}{\kappa^2}
=B^2 -\frac{(B-H_{c2})^2}{{\cal F}'+1}, 
\end{eqnarray}
with 
\begin{eqnarray}
{\cal F}'=\frac{T}{4 \kappa^2 \bar{h}_s^2} \sum_{\omega_n>0}
\sum_m \overline{ 
\langle
ff^{\dagger}(\Delta_m f^\dagger +\Delta_m^{\ast} f)
\rangle_m
}
-\frac{\overline{h_s^2}}{\bar{h}_s^2}. 
\label{eq:Ff}
\end{eqnarray}
Since we consider the case $\kappa \gg 1$ here, 
the last term with $\overline{h_s^2}$ in Eq. (\ref{eq:Ff}) is neglected. 
The free-energy minimum corresponds to 
the minimum of ${\cal F}'$. 

Using $D_i(\epsilon_{\rm F})$ and ${\bf v}_i$ obtained by the DFT calculation 
above, we solve the gap equation (\ref{eq:gapM}) of matrix form 
and obtain $H_{\rm c2}$ and $\Delta_m$ at a given $T$. 
Substituting $\Delta_m$ to Eq. (\ref{eq:fN}) 
we obtain quasi-classical Green's function $f$. 
By these $\Delta_m$ and $f$, we calculate ${\cal F}'$ in Eq. (\ref{eq:Ff}). 
These calculations are performed 
for possible orientations ($0^\circ \le \phi \le 30^\circ$) 
of the triangular vortex lattice, 
and we determine the stable orientation of the vortex lattice 
along the $H_{\rm c2}$ line, by the free-energy minimum.  

 %%%%%%%%%%%%%%%%%%%%%%%%%%%%%%%%%%%%%%%% 
\begin{figure}[tb] 
\begin{center}
\includegraphics[width=8.5cm]{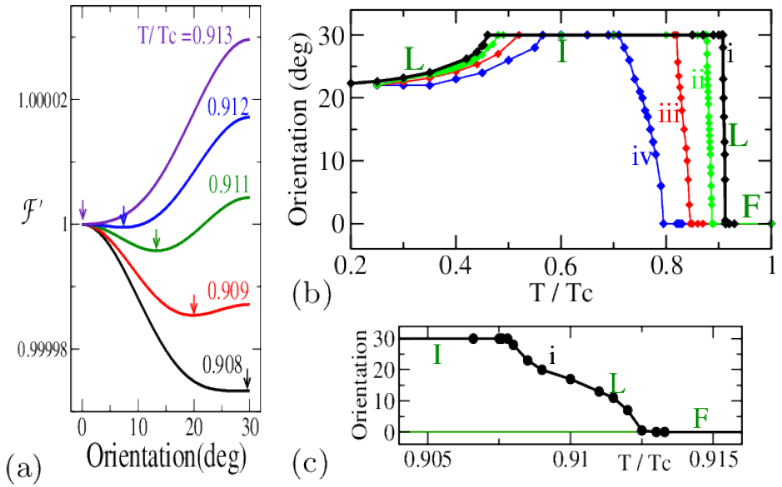} 
\vspace{-0.3cm}
\end{center}
 \caption{ 
(Color online) 
(a)
${\cal F}'(\phi)/{\cal F}'(\phi=0)$ as a function of $\phi$ at 
$T/T_{\rm c}=0.908, 0.909, 0.911, 0.912$, and 0.913, 
showing transition ${\rm I} \rightarrow {\rm L} \rightarrow {\rm F}$, 
for (i) $V_{\sigma,\pi}=0.15V_{\sigma,\sigma}$, $V_{\pi,\pi}=0.32V_{\sigma,\sigma}$. 
The intervals of data points are $1^\circ$ along each line.
Arrows indicate the minimum of each line. 
(b) 
Stable orientation $\phi$ of triangular vortex lattice 
as a function of $T/T_{\rm c}$ 
along $H_{\rm c2}$ line. 
In addition to case (i), we also show lines for 
(ii) $V_{\sigma,\pi}=V_{\sigma,\sigma}/3$, $V_{\pi,\pi}=0$, 
(iii) $V_{\sigma,\pi}=V_{\sigma,\sigma}/2$, $V_{\pi,\pi}=0$, and 
(iv) $V_{\sigma,\pi}=V_{\sigma,\sigma}/3$, $V_{\pi,\pi}=V_{\sigma,\sigma}/2$. 
(c) The high-$T$ range of the transition 
${\rm I} \rightarrow {\rm L} \rightarrow {\rm F}$ of (i) is focused on. 
}
\label{fig:F} 
\end{figure} 
%%%%%%%%%%%%%%%%%%%%%%%%%%%%%%%%%%%%%%%% 

In Fig. \ref{fig:F}(a), we show 
${\cal F}'(\phi)/{\cal F}'(\phi=0)$ as a function of $\phi$ at some $T$ values. 
There, ${\cal F}'$ has a minimum at $\phi=0^\circ$ for $T \ge 0.913$. 
That is, the ${\rm F}$-phase is stable at high-$T$ 
and low-$H$ near $T_{\rm c}$.
For $0.908 < T/T_{\rm c} < 0.913$, 
the minimum orientation $\phi$ changes gradually from $0$ to $30^\circ$, 
indicating that the ${\rm L}$-phase is stable.  
For $0.46 \le T/T_{\rm c} \le 0.908$ at higher fields, 
the ${\rm I}$-phase of $\phi=30^\circ$ is stable. 
The $T$ dependence of the stable orientation $\phi$ is presented 
in Figs. \ref{fig:F}(b) and \ref{fig:F}(c). 
At a low temperature $T/T_{\rm c} < 0.48$ at high fields, 
we see the reentrant transition to the ${\rm L}$-phase 
where $\phi$ decreases from $30^\circ$. 
These successive transitions of the vortex lattice, i.e.,  
${\rm L} \rightarrow {\rm I} \rightarrow {\rm L} \rightarrow {\rm F}$ 
from a low-$T$ along $H_{\rm c2}$, qualitatively reproduce 
the phase diagram observed in the SANS experiment~\cite{Das}.  
Interestingly, we note that 
data points at the lowest $T$ (2 K) and high field (1.8 T) 
show $\phi \sim 23^\circ$~\cite{Das}, which nicely correspond 
to our result of $\phi$ at the lowest $T$ in Fig. \ref{fig:F}(b).    

 To confirm that these qualitative results in Fig. \ref{fig:F}(b) 
do not strongly depend on the interaction parameters 
within appropriate values for ${\rm MgB_2}$,  
we show stable orientations also in other cases 
(ii) $V_{\sigma,\pi}=V_{\sigma,\sigma}/3$, $V_{\pi,\pi}=0$, 
(iii) $V_{\sigma,\pi}=V_{\sigma,\sigma}/2$, $V_{\pi,\pi}=0$, 
and (iv) $V_{\sigma,\pi}=V_{\sigma,\sigma}/3$, $V_{\pi,\pi}=V_{\sigma,\sigma}/2$. 
The gap ratios are $\Delta_\pi/\Delta_\sigma =$0.24(i), 
0.29(ii), 0.40(iii), and 0.57(iv).
We see similar successive transitions there, 
and find that 
the range of the ${\rm F}$-phase (${\rm I}$-phase) 
becomes wider (narrower) 
with the increase in $\Delta_\pi/\Delta_\sigma$. 
This indicates that the stability of the ${\rm F}$-phase is 
due to the contribution of the $\pi$-band. 

\begin{figure}[tb] 
\begin{center}
\includegraphics[width=5.5cm]{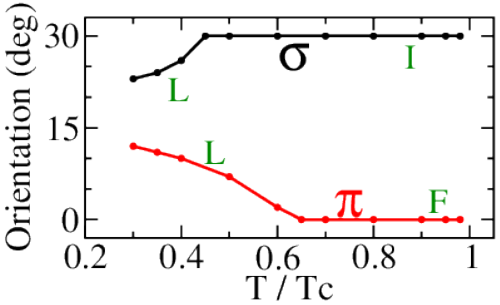} 
\vspace{-0.3cm}
\end{center}
 \caption{ 
(Color online) 
Stable orientation $\phi$ as a function of $T/T_{\rm c}$ 
along $H_{\rm c2}$ line 
when only the $\sigma$-band is superconducting 
($V_{\sigma,\pi}=V_{\pi,\pi}=0$) 
and 
when only the $\pi$-band is superconducting 
($V_{\sigma,\pi}=V_{\sigma,\sigma}=0$).
} 
\label{fig:O2} 
\end{figure}

To clarify the contributions of the two bands, which are competing 
with each other, 
we evaluate the stable orientation of each band. 
As shown in Fig. \ref{fig:O2}, 
if we assume that only the $\sigma$-band is superconducting 
($V_{\sigma,\pi}=V_{\pi,\pi}=0$), 
the stable vortex lattice is the ${\rm I}$-phase with $\phi=30^\circ$ 
at $T \ge 0.45T_{\rm c}$, reflecting the Fermi velocity anisotropy 
in Fig. \ref{fig:DFT}(b). 
Namely, the nearest neighbor of vortices is oriented in the 
direction where $D_\sigma(\varphi_v)$ is small.  
At a lower $T$ and a higher $H$ along $H_{\rm c2}$, 
$\phi$ decreases from $30^\circ$, changing to the ${\rm L}$-phase. 
On the other hand, if we assume that only the $\pi$-band is superconducting 
($V_{\sigma,\pi}=V_{\sigma,\sigma}=0$), 
we obtain opposite results,  
reflecting the opposite Fermi velocity anisotropy in 
Fig. \ref{fig:DFT}(c). 
That is, the stable vortex lattice is the ${\rm F}$-phase 
with $\phi=0^\circ$ at $T \ge 0.65T_{\rm c}$, 
which changes to the ${\rm L}$-phase with increasing $\phi$ 
from $0^\circ$ at a low $T$. 
The reason why the vortex lattice orientation moves at a low $T$ 
along $H_{\rm c2}$ in both cases is 
the Fermi velocity anisotropy changing its contributions at high fields. 
That is, the anisotropy in the $\sigma$ ($\pi$)-band contributes 
to stabilize the ${\rm I}$- (${\rm F}$-) phase at low fields, 
but makes the low-field-phase unstable at high fields. 
This effect is also seen in the case of a square vortex lattice of 
four-fold symmetric superconductors~\cite{Suzuki}. 
To discuss the reason for these behaviors, 
we consider the $\beta$-model of the cylindrical Fermi surface.~\cite{Suzuki}  
If we assume sixfold anisotropy $|{\bf v}_i| \propto 1+0.4\cos 6 \varphi_v$  
for the Fermi velocity in Eq. (\ref{eq:Eil}), 
the stable orientation is at $0^\circ$. 
Instead, if we assume anisotropy 
$D_i(\epsilon_{\rm F}) \propto 1/(1+0.4 \cos 6 \varphi_v)$  
for the Fermi surface DOS, the stable orientation is at $30^\circ$. 
Combining both anisotropies of ${\bf v}_i$ and $D_i(\epsilon_{\rm F})$, 
the stable orientation is at $0^\circ$ at low fields, 
and changes to $30^\circ$ at high fields. 
Since there is usually the relation 
$D_i(\epsilon_{\rm F}) \propto 1/|{\bf v}_i|$, 
the anisotropies of ${\bf v}_i$ and $D_i(\epsilon_v)$ 
tend to compete with each other, 
inducing the change of stable orientation between low fields and high fields.

 %%%%%%%%%%%%%%%%%%%%%%%%%%%%%%%%%%%%%%%% 
%%%%%%%%%%%%%%%%%%%%%%%%%%%%%%%%%%%%%%%% 

On the basis of the results in Fig. \ref{fig:O2}, 
we discuss the origin of the phase diagram of the vortex lattice 
presented in Fig. \ref{fig:F}(b). 
The ${\rm F}$-phase near $T_{\rm c}$ is given 
by the contribution of the $\pi$-band, 
since the total DOS $N_\pi$ and the anisotropy of the $\pi$-band is large 
compared with that of the $\sigma$-band, 
as discussed in Fig. \ref{fig:DFT}. 
However, the contribution of the $\pi$-band decreases at high fields, 
since the superconductivity in the $\pi$-band becomes normal-state-like 
there~\cite{Gonnelli,IchiokaMgB2}. 
This is because the effective upper-critical field of the $\pi$-band is smaller 
than that of the $\sigma$-band, 
which is roughly proportional to the gap amplitude 
$\Delta_\pi/\Delta_\sigma$. 
Thus, since the effective upper-critical field decreases 
for ${\rm (iv) \rightarrow (iii) \rightarrow (ii) \rightarrow (i)}$, 
the $\pi$-band contributions becomes weaker in this order in 
Fig. \ref{fig:F}(b). 
The ${\rm I}$-phase appears at low-$T$ and high fields, 
owing to the contributions of the $\sigma$-band. 
Therefore, the transition ${\rm I} \rightarrow {\rm L} \rightarrow {\rm F}$ 
in the high-$T$ region occurs 
owing to the competition of sixfold anisotropy between the two bands. 
The low-$T$ reentrant behavior from the ${\rm I}$-phase to the ${\rm L}$-phase 
in Fig. \ref{fig:F}(b)  
is due to the anisotropy of the $\sigma$-band. 
The reentrance to the ${\rm L}$-phase occurs 
similarly in the vortex lattice behavior of the $\sigma$-band 
in Fig. \ref{fig:O2}. 

When our results are compared with 
those in experimental observation~\cite{Das},    
it is noted that
the phase boundaries in Fig. \ref{fig:exp} 
are extrapolated near $H_{\rm c2}$. 
The weak anisotropy of the $s$-wave superconducting gap 
and impurity scattering, which are not considered in this study, 
may be factors of minor quantitative modification  
for our theoretical estimation of a stable vortex lattice configuration. 

A recent SANS experiment~\cite{Biswas} 
on ${\rm CaAlSi}$, which has the same ${\rm AlB_2}$-type crystal structure
and a similar two-band feature,~\cite{Kuroiwa}   
has shown a first-order orientation transition of the vortex lattice. 
The present methodology
is readily applicable to this and other omnipresent 
multiband  superconductors, such as Fe pnictides.

In summary, 
a method for the microscopic  Eilenberger analysis of a stable vortex lattice 
configuration along $H_{\rm c2}$ combined with the first-principles
band calculation of DFT is developed for a two-band superconductor. 
We apply this new formulation to ${\rm MgB_2}$, 
and explain the successive transition~\cite{Das} 
${\rm L} \rightarrow {\rm I} \rightarrow {\rm L} \rightarrow {\rm F}$ 
in the orientation of a triangular vortex lattice,
by properly 
considering the Fermi surface anisotropy of the two bands. 
This successful endeavor 
establishes a method using the first-principles band calculation 
to obtain valuable information on the multiband and anisotropy 
of superconductivity from the vortex lattice configuration and morphology. 
We hope that this approach will be applied to other superconductors 
in future studies. 

We thank M.R. Eskildsen for stimulating discussions and 
information on their experimental results.
We also acknowledge useful discussions with P. Miranovi\'{c}, N. Nakai,   
and H.M. Adachi for the formulation and calculation methods. 
This work was supported by KAKENHI No. 21340103.

%%%%%%%%%%%%%%%%%%%%%%%%%%%%%%%%%%%%%%%%%%%%%%%%%%%%%%%%%%%%%%%% 

%%%%%%%%%%%%%%%%%%%%%%%%%%%%%%%%%%%%%%%%%%%%%%%%%%%%%%%%%%%%%%%% 

\begin{thebibliography}{99} 

\bibitem{Suzuki}
K. M. Suzuki, K. Inoue, P. Miranovi\'{c}, M. Ichioka, and K. Machida:   
J. Phys. Soc. Jpn. {\bf 79} (2010) 013702.  
 
\bibitem{Laver}  
M. Laver, E. M. Forgan, S. P. Brown, D. Charalambous, D. Fort, C. Bowell, 
S. Ramos, R. J. Lycett, D. K. Christen, J. Kohlbrecher, C. D. Dewhurst, 
and R. Cubitt: 
Phys. Rev. Lett. {\bf 96} (2006) 167002.  

\bibitem{AdachiNb}
H. M. Adachi, M. Ishikawa, T. Hirano, M. Ichioka, and K. Machida: 
J. Phys. Soc. Jpn. {\bf 80} (2011) 113702.

\bibitem{Takanaka}  
K. Takanaka: 
Prog. Theor. Phys. {\bf 50} (1973) 365.  

\bibitem{Zhitomirsky}
M. E. Zhitomirsky and V.-H. Dao: 
Phys. Rev. B {\bf 69} (2004) 054508.

\bibitem{Kogan} 
V. G. Kogan, P. Miranovi\'{c}, Lj. Dobrosavljevi\'{c}-Gruji\'{c}, 
W. E. Pickett, and D. K. Christen:  
Phys. Rev. Lett. {\bf 79} (1997) 741. 

\bibitem{Gonnelli} 
R. S. Gonnelli, D. Daghero, G. A. Ummarino, V. A. Stepanov, J. Jun, 
S. M. Kazakov, and J. Karpinski: 
Phys. Rev. Lett. {\bf 89} (2002) 247004.

\bibitem{Schmidt} 
H. Schmidt, J. F. Zasadzinski, K. E. Gray, and D. G. Hinks: 
Physica C {\bf 385} (2003) 221. 

\bibitem{Yang}
H. D. Yang, J.-Y. Lin, H. H. Li, F. H. Hsu, C. J. Liu, S.-C. Li, R.-C. Yu, 
and C.-Q. Jin: 
Phys. Rev. Lett. {\bf 87} (2001) 167003. 

\bibitem{Bouquet}
F. Bouquet, R. A. Fisher, N. E. Phillips, D. G. Hinks, and J. D. Jorgensen: 
Phys. Rev. Lett. {\bf 87} (2001) 047001. 

\bibitem{NakaiJPSJ}
N. Nakai, M. Ichioka, and K. Machida: 
J. Phys. Soc. Jpn. {\bf 71} (2002) 23. 

\bibitem{Koshelev}
A. E. Koshelev and A. A. Golubov:  
Phys. Rev. Lett. {\bf 90} (2003) 177002.

\bibitem{IchiokaMgB2}
M. Ichioka, K. Machida, N. Nakai, and P. Miranovi\'{c}: 
Phys. Rev. B {\bf 70} (2004) 144508. 

\bibitem{Moshchalkov}
V. Moshchalkov, M. Menghini, T. Nishio, Q. H. Chen, A. V. Silhanek, 
V. H. Dao, L. F. Chibotaru, N. D. Zhigadlo, and J. Karpinski: 
Phys. Rev. Lett. {\bf 102} (2009) 117001. 

\bibitem{Cubitt}
R. Cubitt, M. R. Eskildsen, C. D. Dewhurst, J. Jun, S. M. Kazakov, 
and J. Karpinski: 
Phys. Rev. Lett. {\bf 91} (2003) 047002.

\bibitem{Das}
P. Das, C. Rastovski, T. R. O'Brien, K. J. Schlesinger, C. D. Dewhurst, 
L. DeBeer-Schmitt, N. D. Zhigadlo, J. Karpinski, and M. R. Eskildsen: 
Phys. Rev. Lett. {\bf 108} (2012) 167001. 

\bibitem{Wien2k} 
P. Blaha, K. Schwarz, G. Madsen, D. Kvasnicka, and
J. Luitz: {\it WIEN2k} (Vienna Univ. Technology, Austria, 2001).
 
\bibitem{Wan}
X. Wan, J. Dong, H. Weng, and D. Y. Xing: 
Phys. Rev. B {\bf 65} (2001) 012502.

\bibitem{Blochl}
P. E. Bl\"{o}chl, O. Jepsen, and O. K. Andersen: 
Phys. Rev. B {\bf 49} (1994) 16223.

\bibitem{Kortus}
J. Kortus, I. I. Mazin, K. D. Belashchenko, V. P. Antropov, and L. L. Boyer: 
Phys. Rev. Lett. {\bf 86} (2001) 4656. 

\bibitem{Eilenberger} 
G. Eilenberger: 
Z. Phys. {\bf 214} (1968) 195. 

\bibitem{Klein} 
U. Klein: 
J. Low Temp. Phys. {\bf 69} (1987) 1. 

\bibitem{Golubov}
A. A. Golubov, J. Kortus, O. V. Dolgov, O. Jepsen, Y. Kong,
O. K. Andersen, B. J. Gibson, K. Ahn, and R. K. Kremer: 
J. Phys.: Condens. Matter {\bf 14} (2002) 1353. 

\bibitem{KitaPRB}
T. Kita and M. Arai: 
Phys. Rev. B {\bf 70} (2004) 224522.   
  
\bibitem{KitaJPSJ}
T. Kita: 
J. Phys. Soc. Jpn. {\bf 67} (1998) 2067.  
 
\bibitem{Nakai} 
N. Nakai, P. Miranovi\'{c}, M. Ichioka, and K. Machida: 
Phys. Rev. Lett. {\bf 89} (2002) 237004.  

\bibitem{Abrikosov}
A. A. Abrikosov: 
Zh. Eksp. Teor. Fiz. {\bf 32} (1957) 1442 
[Sov. Phys. JETP {\bf 5} (1957) 1174]. 
 
\bibitem{deGennes}
P. G. de Gennes:  
{\it Superconductivity of Metals and Alloys} 
(Addison Wesley, CA, 1989) Sec. 6-7. 

\bibitem{Biswas} 
P. K. Biswas, M. R. Lees, G. Balakrishnan, D. Q. Liao, D. S. Keeble, 
J. L. Gavilano, N. Egetenmeyer, C. D. Dewhurst, and D. M. Paul: 
Phys. Rev. Lett. {\bf 108} (2012) 077001. 

\bibitem{Kuroiwa} 
S. Kuroiwa, A. Nakashima, S. Miyahara, N. Furukawa, and J. Akimitsu: 
J. Phys. Soc. Jpn. {\bf 76} (2007) 113705. 


\end{thebibliography}
\end{document}